\documentclass[runningheads]{llncs}
\makeatletter
\renewcommand\subsubsection{\@startsection{subsubsection}{3}{\z@}%
                       {-18\p@ \@plus -4\p@ \@minus -4\p@}%
                       {0.5em \@plus 0.22em \@minus 0.1em}%
                       {\normalfont\normalsize\bfseries\boldmath}}
\makeatother
\usepackage{makecell}

\usepackage{graphicx}
\usepackage{cite}
\usepackage{amsmath,amssymb,amsfonts}
\usepackage{algorithmic}
\usepackage{graphicx}
\usepackage{textcomp}
\usepackage{xcolor}
\usepackage{comment}
\usepackage{hyphenat}
\usepackage{multirow}
\usepackage{tabu}
\usepackage{caption}
\usepackage{subcaption}
\usepackage{
tikz,
relsize,
amsmath,
booktabs,
tikz
}
\usetikzlibrary{arrows,calc,positioning,shadows,shapes}
\usepackage{pgfplots}
\usepackage{pgfplotstable}
\pgfplotsset{compat=1.6}

\usepackage[T1]{fontenc}
\usepackage[utf8]{inputenc}
\usepackage{lmodern}

\usepackage[
labelfont=sf,
hypcap=false,
format=hang,
width=0.8\columnwidth
]{caption}

\usetikzlibrary{
calc,trees,shadows,positioning,arrows,chains,
decorations.pathreplacing,
decorations.pathmorphing,
decorations.shapes,
decorations.text,
shapes,
shapes.geometric,
shapes.symbols,
matrix,
patterns,
intersections,
fit}

\pgfdeclarelayer{background layer}
\pgfdeclarelayer{foreground layer}
\pgfsetlayers{background layer,main,foreground layer}

\tikzset{
>=latex
}
\usepackage[scr=rsfs]{mathalpha}
\usepackage{amsmath}
\usepackage{amssymb}
\usepackage{arydshln}
\usepackage{mathrsfs}
\usepackage[export]{adjustbox}
\usepackage{float}

\usepackage{times}
\usepackage{fancyhdr,graphicx,amsmath,amssymb}
\usepackage[ruled,vlined]{algorithm2e}
\include{pythonlisting}

\def\BibTeX{{\rm B\kern-.05em{\sc i\kern-.025em b}\kern-.08em
    T\kern-.1667em\lower.7ex\hbox{E}\kern-.125emX}}
\begin{document}

%\blindtext
%
%\includepackage[table]{xcolor}
%\title{Beyond information exchange: An approach to render information flow between closest acquaintances}
\title{Beyond Information Exchange: An Approach to Deploy Network Properties for Information Diffusion}
%\title{}

\author{Soumita Das\orcidID{0000-0002-2412-9525} \and
Anupam Biswas\orcidID{0000-0003-0756-6026} \and Ravi Kishore Devarapalli\orcidID{0000-0001-5988-7476}}

\institute{Department of Computer Science and Engineering,\\ National Institute of Technology, Silchar-788010, Assam, India  \\
\email{wingsoffire72@gmail.com}\\
\email{anupam@cse.nits.ac.in}\\
\email{ravikishoredevarapalli@gmail.com}}

\maketitle                      % typeset the header of the contribution
\begin{abstract}

Information diffusion in Online Social Networks is a new and crucial problem in social network analysis field and requires significant research attention. Efficient diffusion of information are of critical importance in diverse situations such as; pandemic prevention, advertising, marketing etc. Although several mathematical models have been developed till date, but previous works lacked systematic analysis and exploration of the influence of neighborhood for information diffusion. In this paper, we have proposed Common Neighborhood Strategy (CNS) algorithm for information diffusion that demonstrates the role of common neighborhood in information propagation throughout the network. The performance of CNS algorithm is evaluated on several real-world datasets in terms of diffusion speed and diffusion outspread and compared with several widely used information diffusion models.  Empirical results show CNS algorithm enables better information diffusion both in terms of diffusion speed and diffusion outspread.

\end{abstract}

\section{Introduction}
\label{introduction}

Online Social Networks (OSNs) plays a key functional role in modern information sharing. Hence, it is used extensively in information diffusion research to examine real-world information diffusion process. As information sharing in such networks occurs through social contacts, hence the underlying network properties plays a significant role in information diffusion. Recently, extensive research has been conducted to understand the role of network properties on the dynamics of information diffusion in OSNs~\cite{goldenberg2001talk,watts2011simple,galstyan2007cascading,peng2020network}. In particular, the propagation of information across these OSNs provides indications of diverse events and situational awareness, such as; marketing prediction, social consciousness, terrorist activities~\cite{bakshy2012role}. It is important to understand how the information will spread throughout the network in the future. Hence, various information diffusion models have been developed to understand the information diffusion behavior in real-world. The information diffusion models are broadly classified into two categories: predictive models and epidemic models. Predictive models considers the network structure to analyze dispersion. For instance, popularity of a book among a population can be predicted by using this model. Whereas, epidemic models considers influence to model epidemiological processes. In particular, the spread of malware are modelled by epidemic models. Several variants  of predictive models and epidemic models have been introduced till date. For example, Independent Cascade (IC) Model, Linear Threshold (LT) model  and Game Theory (GT) model are sender-centric, receiver-centric and profit-centric based predictive models respectively. Whereas, SI (Susceptible Infected) model,  SIR (Susceptible Infected Removed) model,  SIS (Susceptible Infected Susceptible) model and SIRS (Susceptible Infected Removed Susceptible) model are epidemic models~\cite{das2021deployment,li2017,chen2017modeling,bhattacharya2021study}.

Overall, prior work on information diffusion research has either focused on wideness of information diffusion i.e. total number of infected nodes in each iteration or the efficiency of information diffusion i.e. total time required to reach steady state. But the balance between these two important aspects has been underexplored. A good understanding of the information diffusion outspread is important in situations where information needs to be propagated among a wider section of population. For example, Wider spread is required in marketing a product. Whereas,  diffusion speed is important in applications where rate of information flow is to be prioritized. For instance, on election day, a get-out-the-vote campaign is dependent on diffusion speed. In light of these significant applications of information diffusion on real lives, we have proposed CNS algorithm.

\subsection{Basic Idea}
\label{problemformulation}

In Online Social Networks (OSNs), individuals share hundreds or even thousands of connections which associates the individuals to friends, colleagues, family etc. However, all of these connections are not equally strong. Studies suggest that strong connections reside within densely connected nodes~\cite{granovetter1973strength,goldenberg2001talk,centola2007complex,onnela2007structure}. So, identification of these strong connections for information diffusion facilitates initiating information exchange between strongly connected individuals which also triggers exchange of information between the corresponding densely connected neighbors. As information propagation in OSNs expands through social contacts, thus exposure to these multiple sources facilitates rapid diffusion of information throughout the social network. For example, friendship relationship in OSNs may influence a group of users to  download an App or a software that is compatible to his/ her friends to stay in touch. Therefore, investigating and deploying strong connections/ relationships for information diffusion between densely connected neighbors in OSNs is important to analyze it's effectiveness in information propagation throughout the network.

\subsection{Contributions}

In this paper, a novel network property based information diffusion algorithm has been presented. The contributions of this paper are as follows:

\begin{itemize}
    
    \item Utilization of network properties in the diffusion of information throughout the network have been addressed. It demonstrates the importance common neighborhood in information propagation.
    
    \item Illustrates the efficiency of network properties in faster and wider information propagation by evaluating and comparing the proposed algorithm with several popular information diffusion models based on diffusion speed and diffusion outspread.

\end{itemize}

\section{Method}
\label{method}

In this section, we have discussed about our proposed information diffusion algorithm called, Common Neighborhood  Strategy (CNS). Information diffusion in social networks spreads through social contacts which indicates that the underlying network structure plays a significant role in the information diffusion pattern. We have considered  network property namely, common neighborhood to examine it's effect on the dynamics of information diffusion. Before proceeding further, let us first formalize some of the mostly used terms. Suppose, we have a graph $G(V, E)$  where $V$ indicates  set  of  nodes, $E$ indicates  set  of edges. For any connected node pair $(v,u) \in V$, an edge $e_{v,u}$ indicates connection from node $v$ to node $u$, set of neighbors of node $v$ is represented by $\Gamma(v)$.

\subsection{Common Neighborhood}

    For a connected node pair say $(v,u)$, the common neighborhood $\rho_{v,u}$ is defined in terms of neighbors between nodes $u$ and $v$.  Higher common neighborhood score indicates greater interaction frequency and hence, higher similarity. Depending on the number of common neighbors shared by connected node pairs, common neighborhood $\rho_{v,u}$ is computed by considering following cases:\\

    ~~\textbf{Case a.} Nodes $v$ and $u$ do not share any common neighbor i.e. $|\Gamma(v) \cap  \Gamma(u)| = \phi$, then common neighborhood of connected node pair $(v,u)$ is defined by,

\begin{equation}
\label{esz}
\rho_{v,u}  =\left\{\begin{matrix}
1, & if~ \mid \Gamma_{v} \mid ~=1 ~or~\mid \Gamma_{u} \mid=1 \\ 
0, & if ~\mid \Gamma_{v} \mid>1~or~\mid \Gamma_{u} \mid>1
\end{matrix}\right.
\end{equation}
In equation~\ref{esz}, $\rho_{v,u}=1$ if either $\Gamma_{v}=1$ or $\Gamma_{u}=1$ because in this case, $v \leftrightarrow u$ is the only available path for information exchange. Hence, this path will have maximum interaction frequency for any kind of interaction between nodes $v$ and $u$.\\

~~\textbf{Case b.} Nodes $v$ and $u$ shares greater than or equal to one common neighbor i.e. if $|\Gamma(v) \cap  \Gamma(u)| >= 1$, then common neighborhood of connected node pair $(v,u)$ is defined by,

\begin{equation}
\label{eso}
\footnotesize
\resizebox{.994\hsize}{!}{$\rho_{v,u}  =  |\Gamma(v) \cap \Gamma(u)| + |\Gamma(v) \cap \Gamma(z)|  +  |\Gamma(u) \cap \Gamma(z)| +  |\sigma_{vu}| + |\Gamma(w)  \cap \Gamma(z)|,~~  \forall (w,z) \in (\Gamma(v) \cap \Gamma(u)) ~if~  e_{w,z} \in E,  ~w \ne z. $} 
\end{equation}

There are five terms in Equation~\ref{eso}. First term $\mid \Gamma(v) \cap \Gamma(u) \mid$ indicates number of common neighbors shared by nodes $v$ and $u$, second term $\mid \Gamma(v) \cap \Gamma(z) \mid$ and third term $\mid \Gamma(u) \cap \Gamma(z) \mid$ indicates number of common neighbors shared by nodes $v$ and $u$ with common neighbor of $v$ and $u$ respectively, forth term $\mid \sigma_{vu}  \mid$ indicates number of connections shared by common neighbors of $v$ and $u$,  fifth term ~~$\mid \Gamma(w)  \cap \Gamma(z) \mid$ indicates number of common neighbors shared by common neighbors of $v$ and $u$.  
\begin{comment}
Next, after the common neighborhood $\Phi_{v,u}$ have been computed for all the connected node pairs, the common neighborhood is computed by,

\begin{equation}
    \label{info}
    \rho_{v,u} =  \frac{\Phi_{v,u}}{max_{u \in \Gamma_{v}}\Phi_{v,u}}
\end{equation}
In equation~\ref{info}, the term in the denominator indicates  maximum common neighborhood score shared by $v$ with it's neighboring node. If $\Phi_{v,u}= 0$, then  $\rho_{v,u}=0$.\\

\end{comment}

 \textbf{Neighborhood Density:} For a connected node pair $(v,u)$, neighborhood density is used to compute the density of it's closely connected neighbors. This density score is utilized to measure the tie strength of a connected node pair. Greater tie strength score indicates higher proximity. For any connected node pair $(v,u)$, neighborhood density/ tie strength is computed by using  common neighborhood $\rho_{v,u}$ and is defined by,
 
 \begin{equation}
    \label{info}
    \Phi_{v,u} =  \frac{\rho_{v,u}}{max_{u \in \Gamma_{v}}\rho_{v,u}}
\end{equation}

In equation~\ref{info}, the term in the denominator indicates  maximum common neighborhood score shared by $v$ with it's neighboring node. The denominator term makes tie strength $ \Phi_{v,u}$ asymmetric.  If $\Phi_{v,u}=1$, it indicates that the tie strength shared by nodes $v$ and $u$ is maximum. Whereas,  if $\rho_{v,u}= 0$, then  $\Phi_{v,u}=0$, which means that if the common neighborhood of a node pair is 0, then the tie strength is 0.  In this context, a tie between node pair $(v,u)$ is represented by edge $e_{v,u} \in E$  indicating connection from node $v$ to node $u$.  \(\mathscr{U}\) contains list of edges in graph $G$ having maximum tie strength score. \\

\subsection{Common Neighborhood Strategy}

Here, we present the Common Neighborhood Strategy (CNS) algorithm. This algorithm has been developed to investigate the effectiveness of common neighborhood for information diffusion. The proposed CNS algorithm comprises of three aspects. Let us consider that the diffusion process starts from a node $v \in V$. The first aspect is to identify the adjacent node/ nodes of $v$ with which it shares maximum tie strength i.e. if $\Phi_{v,u}=1$, it indicates that node $v$ shares maximum tie strength with node $u$. Then, information diffusion is initiated from node $v$ to node $u$ and is defined by,\\

\begin{equation}
 \label{first}
    \tau_{1}=  \sum_{\substack{u \in \Gamma(v), ~e_{v,u} \in  \mathscr{U}}}\phi_{v \rightarrow u}
\end{equation}

Equation~\ref{first} indicates that node $v$  exchanges information with neighboring node/ nodes, where $\Phi_{v,u}=1$, $\forall u \in \Gamma(v)$. Here, $\phi_{v \rightarrow u}$ indicates that information exchange takes place from node $v$ to node $u$. After identification of node $u$, which shares maximum tie strength with $v$, the second aspect is to exchange information with nodes contributing in the common neighborhood $\rho_{v,u}$ of connected node pair $(v,u)$. Let us assume that $\mathscr{F}$ contains list of nodes contributing in the common neighborhood of node pair $(v,u)$. Then, the second aspect is defined by,\\

\begin{equation}
    \label{second}
     \tau_{2}= \sum_{\substack{i,j \in \mathscr{F}, ~i,j \in V, \\ i \in \Gamma(v), j \in \Gamma(u)}}(\phi_{v \rightarrow i}+ \phi_{u \rightarrow j}) 
\end{equation}

Nodes $v$ and $u$ exchanges information with it's neighbors which contributes in the common neighborhood of connected node pair $(v,u)$ and is indicated by terms $\phi_{v \rightarrow i}$ and $\phi_{u \rightarrow j}$  respectively in equation~\ref{second}. Then, the third aspect is to identify those neighboring nodes of $v$ which have not yet participated in the information diffusion process and which shares maximum tie strength with $v$ i.e. $e_{v,z} \notin  \mathscr{U}$ but $e_{z,v} \in \mathscr{U}$ where $z \in \Gamma(v)$ and is defined by,\\

\begin{equation}
    \label{third}
     \tau_{3}= \phi_{v \rightarrow z},  \forall z \in \Gamma(v), e_{z,v} \in  \mathscr{U}
\end{equation}

\begin{figure}[!ht]
     \centering
     \begin{subfigure}[b]{0.2\textwidth}
         \centering
         \includegraphics[width=\textwidth]{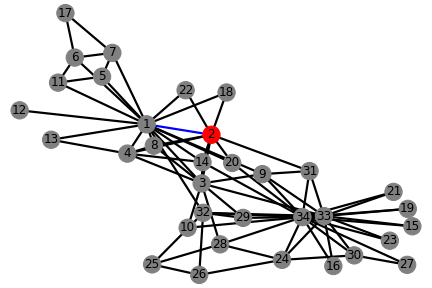}
         \caption{\tiny Step 1: Iteration one}
         \label{two}
     \end{subfigure}
     \hfill
     \begin{subfigure}[b]{0.2\textwidth}
         \centering
         \includegraphics[width=\textwidth]{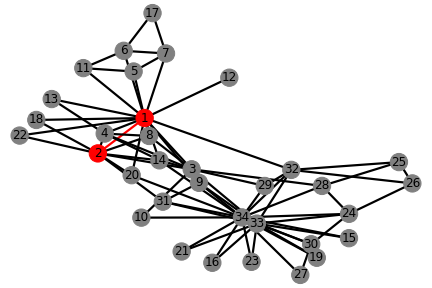}
         \caption{\tiny  Step 2: Iteration one}
         \label{four}
     \end{subfigure}
     \hfill
     \begin{subfigure}[b]{0.2\textwidth}
         \centering
         \includegraphics[width=\textwidth]{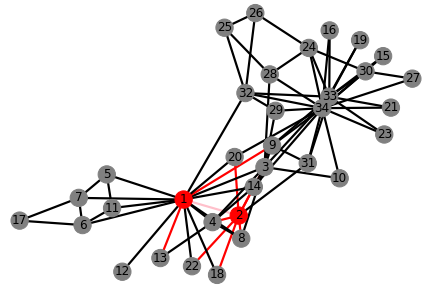}
         \caption{\tiny Step 3: Iteration one}
         \label{five}
     \end{subfigure}
     \hfill
     \begin{subfigure}[b]{0.2\textwidth}
         \centering
         \includegraphics[width=\textwidth]{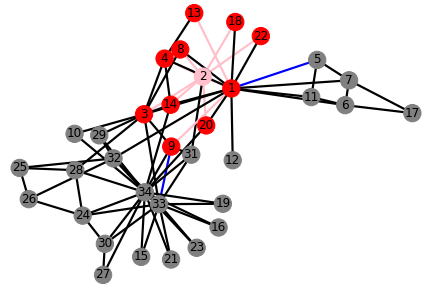}
         \caption{\tiny Step 4: Iteration one}
         \label{six}
     \end{subfigure}
     \hfill
     \begin{subfigure}[b]{0.2\textwidth}
         \centering
         \includegraphics[width=\textwidth]{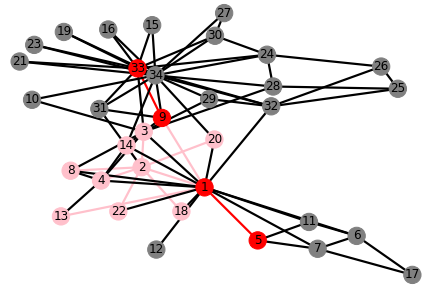}
         \caption{\tiny  Step 5: Iteration two}
         \label{eight}
     \end{subfigure}
      \hfill
     \begin{subfigure}[b]{0.2\textwidth}
         \centering
         \includegraphics[width=\textwidth]{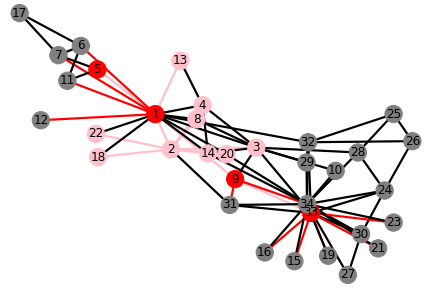}
         \caption{\tiny Step 6: Iteration two}
         \label{nine}
     \end{subfigure}
     \hfill
     \begin{subfigure}[b]{0.2\textwidth}
         \centering
         \includegraphics[width=\textwidth]{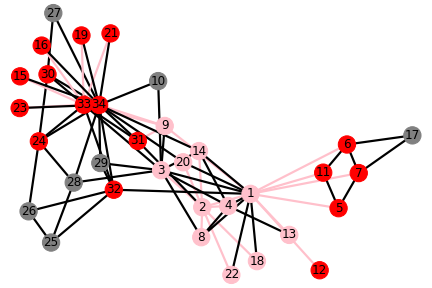}
         \caption{\tiny Step 7: Iteration two}
         \label{ten}
     \end{subfigure}
     \hfill
     \begin{subfigure}[b]{0.2\textwidth}
         \centering
         \includegraphics[width=\textwidth]{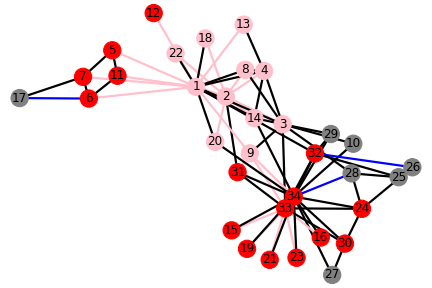}
         \caption{\tiny Step 8: Iteration two}
         \label{twelve}
     \end{subfigure}
     \hfill
       \begin{subfigure}[b]{0.2\textwidth}
         \centering
         \includegraphics[width=\textwidth]{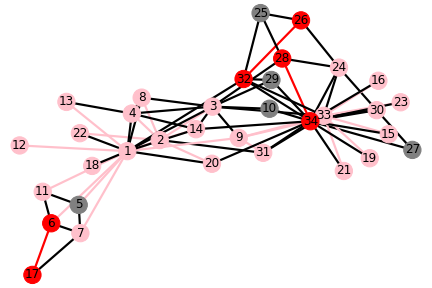}
         \caption{\tiny Step 9: Iteration three}
         \label{thirteen}
     \end{subfigure}
     \hfill
     \begin{subfigure}[b]{0.2\textwidth}
         \centering
         \includegraphics[width=\textwidth]{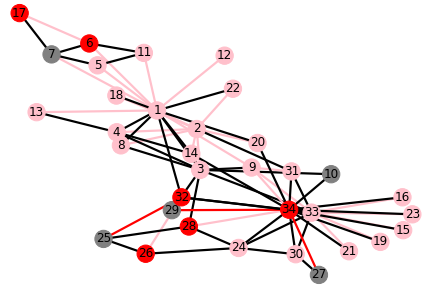}
         \caption{\tiny Step 10: Iteration three}
         \label{fifteen}
     \end{subfigure}
     \hfill
      \begin{subfigure}[b]{0.2\textwidth}
         \centering
         \includegraphics[width=\textwidth]{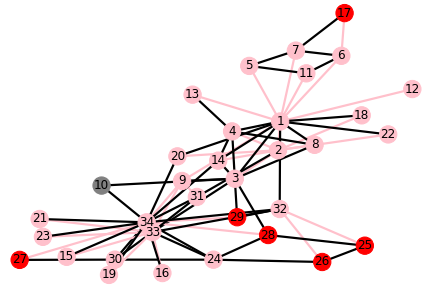}
         \caption{\tiny Step 11: Iteration three}
         \label{thirteen}
     \end{subfigure}
     \hfill
     \begin{subfigure}[b]{0.2\textwidth}
         \centering
         \includegraphics[width=\textwidth]{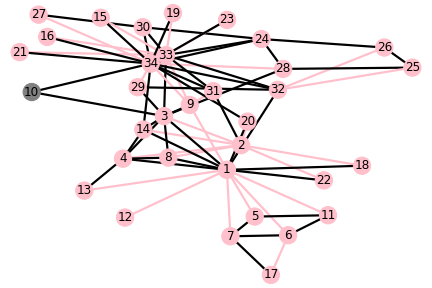}
         \caption{\tiny Step 12: Iteration three}
         \label{sixteen}
     \end{subfigure}
     \hfill
     \caption{\footnotesize Demonstration of CNS algorithm using Karate network. Red colored nodes indicate active nodes, while grey colored nodes indicate inactive nodes, blue arcs refers to strong ties, red arcs represents connections where activated nodes try to activate their inactive neighbors, peach arcs represent edges propagated once and peach nodes indicate nodes already participated in information diffusion.}
      \label{demonstration}
      \end{figure}

Equation~\ref{third} indicates that node $z$ shares maximum tie strength score with node $v$, if node $z$ have not participated in information exchange before, then node $v$ exchanges information with node $z$. Ultimately, combination of the first three aspects presented in equation~\ref{first}, equation~\ref{second} and equation~\ref{third} gives the information diffusion score adopted by CNS algorithm. Information diffusion from node $v$ as defined by CNS  algorithm is given by,\\

\begin{equation}
    \label{forth}
      \mathscr{I}_{v} = \tau_{1} + \tau_{2} + \tau_{3}
\end{equation}
Equation ~\ref{forth} indicates that information diffusion from node $v$ is a combination of $\tau_{1},\tau_{2}$ and $\tau_{3}$. 
The proposed information diffusion based CNS algorithm is designed considering common neighborhood concept. Step-by-step illustration of the diffusion process of CNS algorithm is presented in  Fig.~\ref{demonstration}. Let us assume that node 2 initiates the diffusion process. In Step 1, node 2 identifies the adjacent edge having maximum tie strength score. Here, node 2 shares maximum tie strength with node 1 indicated by blue arc in subfigure (a). Then, in Step 2, node 2 activates node 1. In Step 3, all the nodes contributing to the maximum tie strength score of node pair (2,1) are targeted for activation. At the end of iteration one, nodes 1 and 9 identifies the adjacent edges having maximum tie strength score i.e. arcs (1,5) and (9,33) respectively. Thereafter, in iteration two and three all the steps followed in iteration one are repeated for activating inactive neighbors and ultimately, the diffusion pattern obtained by incorporation of CNS algorithm is shown in subfigure (l) of ~\ref{demonstration} by peach arcs. Pseudocode of CNS method is shown in Algorithm~ 1. \\

\begin{algorithm}
\label{CNSalgo}
 \caption{CNS}
 \begin{algorithmic}[1]
 \renewcommand{\algorithmicrequire}{\textbf{Input:}}
 \renewcommand{\algorithmicensure}{\textbf{Output:}}
 \REQUIRE Social Network, $G=(V,E)$
 \ENSURE  $ $  \tcp{List of activated nodes $ \mathscr{A}$}
 \STATE CNS (G)
  \FOR {each node $v$ in $G$}
     \IF {$e_{v,u}$ in $ \mathscr{U}$}
        \STATE $ $ \tcp{Compute Information Diffusion using equation (7)}
     \ENDIF
   \ENDFOR
   \RETURN  $ \mathscr{A}$   
 \end{algorithmic}
 \end{algorithm}  

\section{Experimental Analysis}
\label{experimentalanalysis}

Evaluation of information diffusion result is necessary to determine the performance of information diffusion models/ algorithms. In this section, we brief the evaluation strategies such as, information diffusion models along with their parameters, diffusion speed evaluation, diffusion outspread evaluation and datasets.  We have considered Independent Cascade  (IC) model with edge threshold set as 1 and Susceptible Infected (SI) model with infection probability set as 0.50 for comparative analysis purpose. IC model has been utilized to examine the role of tie strength in information diffusion behavior. Next, SI model have been considered to investigate the role of influence in information propagation in OSNs. These two models are selected for comparative analysis because tie strength and influence are positively correlated and hence, crucial for examining diffusion performance.

\begin{table}[!t]
\caption {\footnotesize Real-world datasets used in the experiments,  $K$ indicates average degree.}
\label{table1}
\centering
  \begin{tabular} {| m{6.8em}| m{3.6em} | m{3.6em} |m{3.6em}| m{12.4em}|}
   \hline
   \centering{Name} &  \centering{$\mid V \mid$} & \centering{$\mid E \mid$} & \centering{$K$} & ~~~~~~~Network Description \\
  
   \hline
   \centering{Karate}  &\centering{34} & \centering{78} & \centering{4.59} &  \centering{Zachary's Karate Club\cite{zachary1977information}}
   \tabularnewline
     \centering{Les Mis{\'e}rables}  &\centering{77} & \centering{254} & \centering{6.60} & \centering{Les Mis{\'e}rables network\cite{knuth1993stanford}}
   \tabularnewline
   \centering{Jazz}  &\centering{198} & \centering{2742} & \centering{27.70} & \centering{Jazz musician network \cite{gleiser2003community}}   
   \tabularnewline
   \centering{Polblogs}  &\centering{1224} & \centering{16718} & \centering{27.31}  & \centering{Polblogs network~\cite{adamic2005political}} 
   \tabularnewline
    \hline
\end{tabular}
\end{table}

\pgfplotstableread[row sep=\\,col sep=&]{
Number of Iterations & CNS & IC & SI\\
Karate & 3  & 3  &	5 \\
Les Mis{\'e}rables & 3 & 4 & 5 \\
Jazz & 4 & 5 & 6 \\
Polblogs & 4 & 6 & 10 \\
}\lookm

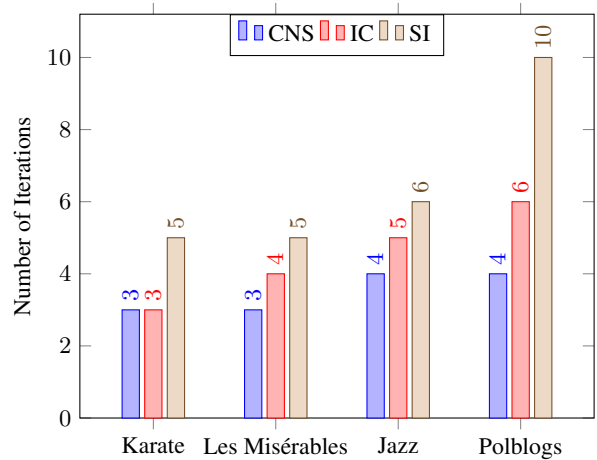
\begin{figure}[!hb]
\centering
\begin{tikzpicture}
    \begin{axis}[
            ybar,
           bar width=.23cm,
            width=0.69\textwidth,
            height=0.57\textwidth,
            enlarge x limits=0.2,
            legend style={at={(0.5,1)},
                anchor=north,legend columns=3,legend cell align=left},
            symbolic x coords={Karate,Les Mis{\'e}rables,Jazz, Polblogs},
            xtick=data,
             x tick label style={rotate=00,anchor=north},
            %nodes near coords,
            nodes near coords align={vertical},
            ymin=0,ymax=11.2,
            ylabel={Number of Iterations},
            nodes near coords,
            every node near coord/.append style={rotate=90, anchor=west}
        ]
        \addplot table[x=Number of Iterations,y=CNS]{\lookm};
        \addplot table[x=Number of Iterations,y=IC]{\lookm};
        \addplot table[x=Number of Iterations,y=SI]{\lookm};
        \legend{CNS, IC, SI}
  ,  \end{axis}
\end{tikzpicture}
\caption{ Number of iterations required by each of the algorithms for information diffusion.}
\label{fig2}
\end{figure}

\usetikzlibrary{positioning}
\tikzset{main node/.style={circle,fill=white,draw,minimum size=.2cm,inner sep=0pt},}
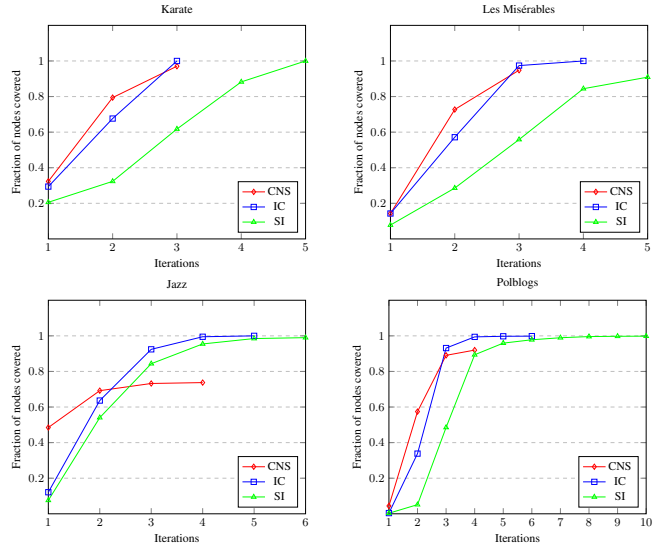
\begin{figure}[!t]
\centering
\begin{subfigure}[b]{0.35\linewidth}
\centering
\usetikzlibrary{quotes}
\begin{tikzpicture}[scale=0.4990]
\begin{axis}[
    title={Karate},%Iterations Vs Fraction of nodes covered},
    xlabel={Iterations},
    ylabel={Fraction of nodes covered},
    xmin=1, xmax=5,
    ymin=0, ymax=1.2,
    xtick={1,2, 3, 4,5},
    ytick={0.20,0.40,0.60,0.80,1.00},
    legend pos=south east,
    ymajorgrids=true,
    grid style=dashed,
]

\addplot[
    color=red,
    mark=diamond,
    ]
    coordinates {
    (1,0.3235)(2,0.7941)(3,0.9705)
    %(10000,2.96)(19990,2.9)
    };
    \addlegendentry{\footnotesize CNS}
\addplot[
    color=blue,
    mark=square,
    ]
    coordinates {
    (1,0.2941)(2,0.6764)(3,1.0)
   
    %(10000,2.9)(19990,2.5)
    };
    \addlegendentry{\footnotesize IC}
\addplot[
    color=green,
    mark=triangle,
    ]
    coordinates {
    
    (1,0.2058)(2,0.3235)(3,0.6176)(4,0.8823)(5,1.0)
    };
    \addlegendentry{\footnotesize SI}
\end{axis}
\end{tikzpicture}
% \includegraphics{}
%\caption{\tiny Karate dataset}%Fraction of nodes covered per iteration on Karate dataset}
%\label{frackarate}
    \end{subfigure}
     \hspace{0.12cm}
    \begin{subfigure}[b]{0.35\linewidth}
\centering
\usetikzlibrary{quotes}
\begin{tikzpicture}[scale=0.4990]
\begin{axis}[
    title={Les Mis{\'e}rables},%Iterations Vs Fraction of nodes covered},
    xlabel={Iterations},
    ylabel={Fraction of nodes covered},
    xmin=1, xmax=5,
    ymin=0, ymax=1.2,
    xtick={1,2, 3, 4,5},
    ytick={0.20,0.40,0.60,0.80,1.00},
    legend pos=south east,
    ymajorgrids=true,
    grid style=dashed,
]

\addplot[
    color=red,
    mark=diamond,
    ]
    coordinates {
    (1,0.1428)(2,0.7272)(3,0.9480)
    %(10000,2.96)(19990,2.9)
    };
    \addlegendentry{\footnotesize CNS}
\addplot[
    color=blue,
    mark=square,
    ]
    coordinates {
    (1,0.1428)(2,0.5714)(3,0.9740)(4,1.0)
   
    %(10000,2.9)(19990,2.5)
    };
    \addlegendentry{\footnotesize IC}
\addplot[
    color=green,
    mark=triangle,
    ]
    coordinates {
    
    (1,0.0779)(2,0.2857)(3,0.5584)(4,0.8441)(5,0.9090)
    };
    \addlegendentry{\footnotesize SI}
\end{axis}
\end{tikzpicture}
% \includegraphics{}
%\caption{\tiny Les Mis{\'e}rables dataset}%Fraction of nodes covered per iteration on Les Mis{\'e}rables dataset}
%\label{fraclesmiserables}
\end{subfigure}  
\hspace{0.12cm}
    \begin{subfigure}[b]{0.35\linewidth}
\centering
\usetikzlibrary{quotes}
\begin{tikzpicture}[scale=0.4990]
\begin{axis}[
    title={Jazz},%Iterations Vs Fraction of nodes covered},
    xlabel={Iterations},
    ylabel={Fraction of nodes covered},
    xmin=1, xmax=6,
    ymin=0, ymax=1.2,
    xtick={1,2, 3, 4,5,6},
    ytick={0.20,0.40,0.60,0.80,1.00},
    legend pos=south east,
    ymajorgrids=true,
    grid style=dashed,
]

\addplot[
    color=red,
    mark=diamond,
    ]
    coordinates {
    (1,0.4848)(2,0.6919)(3,0.7323)(4,0.7373)
    %(10000,2.96)(19990,2.9)
    };
    \addlegendentry{\footnotesize CNS}
\addplot[
    color=blue,
    mark=square,
    ]
    coordinates {
    (1,0.1212)(2,0.6363)(3,0.9242)(4,0.9949)(5,1.0)
   
    %(10000,2.9)(19990,2.5)
    };
    \addlegendentry{\footnotesize IC}
\addplot[
    color=green,
    mark=triangle,
    ]
    coordinates {
    
    (1,0.0757)(2,0.5404)(3,0.8434)(4,0.9545)(5,0.9848)(6,0.9898)
    };
    \addlegendentry{\footnotesize SI}
\end{axis}
\end{tikzpicture}
% \includegraphics{}
%\caption{\tiny Jazz dataset}%Fraction of nodes covered per iteration on Les Mis{\'e}rables dataset}
%\label{fracclesmiserables}
\end{subfigure}
\hspace{0.12cm}
    \begin{subfigure}[b]{0.35\linewidth}
\centering
\usetikzlibrary{quotes}
\begin{tikzpicture}[scale=0.4990]
\begin{axis}[
    title={ Polblogs},%Iterations Vs Fraction of nodes covered},
    xlabel={Iterations},
    ylabel={Fraction of nodes covered},
    xmin=1, xmax=10,
    ymin=0, ymax=1.2,
    xtick={1,2, 3, 4,5,6,7,8,9,10},
    ytick={0.20,0.40,0.60,0.80,1.00},
    legend pos=south east,
    ymajorgrids=true,
    grid style=dashed,
]

\addplot[
    color=red,
    mark=diamond,
    ]
    coordinates {
    (1,0.0441)(2,0.5743)(3,0.8905)(4,0.9199)
    %(10000,2.96)(19990,2.9)
    };
    \addlegendentry{\footnotesize CNS}
\addplot[
    color=blue,
    mark=square,
    ]
    coordinates {
    (1,0.0040)(2,0.3382)(3,0.9313)(4,0.9942)(5,0.9975)(6,0.9983)
   
    %(10000,2.9)(19990,2.5)
    };
    \addlegendentry{\footnotesize IC}
\addplot[
    color=green,
    mark=triangle,
    ]
    coordinates {
    (1,0.0032)(2,0.0514)(3,0.4852)(4,0.8937)(5,0.9591)(6,0.9779)(7,0.9893)(8,0.9959)(9,0.9975)(10,0.9983)
    };
    \addlegendentry{\footnotesize SI}
\end{axis}
\end{tikzpicture}
% \includegraphics{}
%\caption{\tiny Polblogs dataset}%Fraction of nodes covered per iteration on Les Mis{\'e}rables dataset}
%\label{fraccclesmiserables}
\end{subfigure}  
\caption{Fraction of nodes covered per iteration.} %obtained by each algorithm on real-world datasets. Higher scores indicate maximum information diffusion speed. }
\label{fig3}    
\end{figure}
%%%%%%%%%%%%%%%%%%%%%%%%%%%%%%%%%%%%%%%%%%%%%%%%%%%%%%%%%%%%%%%%%%%%%

%%%%%%%%%%%%%%%%%%%%%%%%%%%%%%%%%%%%%

%%%%%%%%%%%%%%%%%%%%%%%%%%%%%%%%%%%%%%%%%%%%%%%%%%%%%%%%%%%%%%%%%%%%%%
\usetikzlibrary{positioning}
\tikzset{main node/.style={circle,fill=white,draw,minimum size=.2cm,inner sep=0pt},}
\begin{figure}[!h]
\centering
\begin{subfigure}[b]{0.35\linewidth}
\centering
\usetikzlibrary{quotes}
\begin{tikzpicture}[scale=0.4990]
\begin{axis}[
    title={Karate},%Iterations Vs Diameter},
    xlabel={Iterations},
    ylabel={Diameter},
    xmin=1, xmax=5,
    ymin=0, ymax=6,
    xtick={1,2, 3, 4,5},
    ytick={1,2,3,4,5},
    legend pos=south east,
    ymajorgrids=true,
    grid style=dashed,
]

\addplot[
    color=red,
    mark=diamond,
    ]
    coordinates {
    (1,2)(2,4)(3,5)
    %(10000,2.96)(19990,2.9)
    };
    \addlegendentry{\footnotesize CNS}
\addplot[
    color=blue,
    mark=square,
    ]
    coordinates {
   (1,2)(2,3)(3,5)
   
    %(10000,2.9)(19990,2.5)
    };
    \addlegendentry{\footnotesize IC}
\addplot[
    color=green,
    mark=triangle,
    ]
    coordinates {
    
    (1,2)(2,3)(3,4)(4,4)(5,5)
    %(15000,5.4)
    };
    \addlegendentry{\footnotesize SI}
\end{axis}
\end{tikzpicture}
%\caption{\tiny Karate dataset }%Distribution of diameter with respect to number of iterations on}
%\label{diameterkarate}
    \end{subfigure}
     \hspace{0.12cm}
     \begin{subfigure}[b]{0.35\linewidth}
\centering
\usetikzlibrary{quotes}
\begin{tikzpicture}[scale=0.499]
\begin{axis}[
    title={Les Mis{\'e}rables},%Iterations Vs Diameter},
    xlabel={Iterations},
    ylabel={Diameter},
    xmin=1, xmax=5,
    ymin=0, ymax=6,
    xtick={1,2, 3, 4,5},
    ytick={1,2,3,4,5},
    legend pos=south east,
    ymajorgrids=true,
    grid style=dashed,
]

\addplot[
    color=red,
    mark=diamond,
    ]
    coordinates {
    (1,2)(2,4)(3,5)
    %(10000,2.96)(19990,2.9)
    };
    \addlegendentry{\footnotesize CNS}
\addplot[
    color=blue,
    mark=square,
    ]
    coordinates {
   (1,2)(2,3)(3,4)(4,5)
   
    %(10000,2.9)(19990,2.5)
    };
    \addlegendentry{\footnotesize IC}
\addplot[
    color=green,
    mark=triangle,
    ]
    coordinates {
    
    (1,2)(2,3)(3,4)(4,4)(5,4)
    %(15000,5.4)
    };
    \addlegendentry{\footnotesize SI}
\end{axis}
\end{tikzpicture}
% \includegraphics{}
%\caption{\tiny Les Mis{\'e}rables dataset}%Distribution of diameter with respect to number of iterations on Les Mis{\'e}rables dataset}
%\label{diameterLESMISERABLES}
\end{subfigure} 
   \hspace{0.12cm}
     \begin{subfigure}[b]{0.35\linewidth}
\centering
\usetikzlibrary{quotes}
\begin{tikzpicture}[scale=0.4990]
\begin{axis}[
    title={Jazz},%Iterations Vs Diameter},
    xlabel={Iterations},
    ylabel={Diameter},
    xmin=1, xmax=6,
    ymin=0, ymax=6,
    xtick={1,2, 3, 4,5,6},
    ytick={1,2,3,4,5},
    legend pos=south east,
    ymajorgrids=true,
    grid style=dashed,
]

\addplot[
    color=red,
    mark=diamond,
    ]
    coordinates {
    (1,3)(2,4)(3,5)(4,5)
    %(10000,2.96)(19990,2.9)
    };
    \addlegendentry{\footnotesize CNS}
\addplot[
    color=blue,
    mark=square,
    ]
    coordinates {
   (1,2)(2,3)(3,4)(4,5)(5,6)
   
    %(10000,2.9)(19990,2.5)
    };
    \addlegendentry{\footnotesize IC}
\addplot[
    color=green,
    mark=triangle,
    ]
    coordinates {
    
    (1,2)(2,3)(3,4)(4,5)(5,5)(6,5)
    %(15000,5.4)
    };
    \addlegendentry{\footnotesize SI}
\end{axis}
\end{tikzpicture}
% \includegraphics{}
%\caption{\tiny Jazz dataset}%Distribution of diameter with respect to number of iterations on Les Mis{\'e}rables dataset}
%\label{diameterrrLESMISERABLES}
\end{subfigure} 
   \hspace{0.12cm}
     \begin{subfigure}[b]{0.35\linewidth}
\centering
\usetikzlibrary{quotes}
\begin{tikzpicture}[scale=0.4990]
\begin{axis}[
    title={Polblogs},%Iterations Vs Diameter},
    xlabel={Iterations},
    ylabel={Diameter},
    xmin=1, xmax=10,
    ymin=0, ymax=9,
    xtick={1,2, 3, 4,5,6,7,8,9,10},
    ytick={1,2,3,4,5,6,7,8},
    legend pos=south east,
    ymajorgrids=true,
    grid style=dashed,
]
\addplot[
    color=red,
    mark=diamond,
    ]
    coordinates {
    (1,2)(2,4)(3,6)(4,6)
    %(10000,2.96)(19990,2.9)
    };
    \addlegendentry{\footnotesize CNS}
\addplot[
    color=blue,
    mark=square,
    ]
    coordinates {
   (1,2)(2,4)(3,5)(4,6)(5,7)(6,8)
   
    %(10000,2.9)(19990,2.5)
    };
    \addlegendentry{\footnotesize IC}
\addplot[
    color=green,
    mark=triangle,
    ]
    coordinates {
    
    (1,2)(2,3)(3,4)(4,5)(5,6)(6,7)(7,7)(8,7)(9,7)(10,8)
    %(15000,5.4)
    };
    \addlegendentry{\footnotesize SI}
\end{axis}
\end{tikzpicture}
% \includegraphics{}
%\caption{\tiny Polblogs dataset}%Distribution of diameter with respect to number of iterations on Les Mis{\'e}rables dataset}
%\label{diameterrLESMISERABLES}
\end{subfigure} 
\caption{ Diameter of diffusion horizon per iteration.}% on real-world datasets. Higher scores indicate maximum information spread per iteration.}
\label{fig4}    
\end{figure}
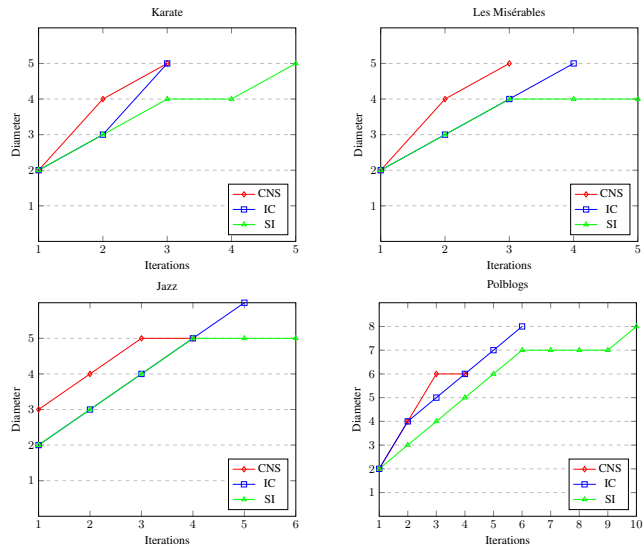

\usetikzlibrary{positioning}
\tikzset{main node/.style={circle,fill=white,draw,minimum size=.2cm,inner sep=0pt},}
\begin{figure}[!t]
\centering
\begin{subfigure}[b]{0.35\linewidth}
\centering
\usetikzlibrary{quotes}
\begin{tikzpicture}[scale=0.499]
\begin{axis}[
    title={Karate},%Iterations Vs Average Distance},
    xlabel={Iterations},
    ylabel={Average Distance},
    xmin=1, xmax=5,
    ymin=0, ymax=3,
    xtick={1,2, 3, 4,5},
    ytick={1.0,1.25,1.5,1.75,2.0,2.25,2.50},
    legend pos=south east,
    ymajorgrids=true,
    grid style=dashed,
]

\addplot[
    color=red,
    mark=diamond,
    ]
    coordinates {
    (1,1.5636)(2,2.2905)(3,2.414)
    %(10000,2.96)(19990,2.9)
    };
    \addlegendentry{\footnotesize CNS}
\addplot[
    color=blue,
    mark=square,
    ]
    coordinates {
    (1,1.5333)(2,1.9762)(3,2.4081)
   
    %(10000,2.9)(19990,2.5)
    };
    \addlegendentry{\footnotesize IC}
\addplot[
    color=green,
    mark=triangle,
    ]
    coordinates {
    
    (1,1.6190)(2,1.6545)(3,2.0333)(4,2.2758)(5,2.4081)
    %(15000,5.4)
    };
    \addlegendentry{\footnotesize SI}
\end{axis}
\end{tikzpicture}
% \includegraphics{}
%\caption{\tiny  Karate dataset}%Distribution of average distance with respect to number of iterations on Karate dataset}
%\label{avgdistancekarate}
    \end{subfigure}
     \hspace{0.12cm}
     \begin{subfigure}[b]{0.35\linewidth}
\centering
\usetikzlibrary{quotes}
\begin{tikzpicture}[scale=0.499]
\begin{axis}[
    title={Les Mis{\'e}rables},%Iterations Vs Average Distance},
    xlabel={Iterations},
    ylabel={Average Distance},
    xmin=1, xmax=5,
    ymin=0, ymax=3,
    xtick={1,2, 3, 4,5},
    ytick={1.0,1.25,1.5,1.75,2.0,2.25,2.50,3.0},
    legend pos=south east,
    ymajorgrids=true,
    grid style=dashed,
]

\addplot[
    color=red,
    mark=diamond,
    ]
    coordinates {
    (1,1.76)(2,2.3188)(3,2.5742)
    %(10000,2.96)(19990,2.9)
    };
    \addlegendentry{\footnotesize CNS}
\addplot[
    color=blue,
    mark=square,
    ]
    coordinates {
    (1,1.7636)(2,2.1183)(3,2.5834)(4,2.6411)
   
    %(10000,2.9)(19990,2.5)
    };
    \addlegendentry{\footnotesize IC}
\addplot[
    color=green,
    mark=triangle,
    ]
    coordinates {
    
    (1,1.6)(2,2.1731)(3,2.3122)(4,2.4841)(5,2.5457)
    %(15000,5.4)
    };
    \addlegendentry{\footnotesize SI}
\end{axis}
\end{tikzpicture}
% \includegraphics{}
%\caption{\tiny Les Mis{\'e}rables dataset}%Distribution of average distance with respect to number of iterations on Les Mis{\'e}rables dataset}
%\label{avgdistancelesmiserables}
\end{subfigure} 
 \hspace{0.12cm}
     \begin{subfigure}[b]{0.35\linewidth}
\centering
\usetikzlibrary{quotes}
\begin{tikzpicture}[scale=0.499]
\begin{axis}[
    title={Jazz},%Iterations Vs Average Distance},
    xlabel={Iterations},
    ylabel={Average Distance},
    xmin=1, xmax=6,
    ymin=0, ymax=2.9,
    xtick={1,2, 3, 4,5,6},
    ytick={1.0,1.25,1.5,1.75,2.0,2.25,2.50},
    legend pos=south east,
    ymajorgrids=true,
    grid style=dashed,
]

\addplot[
    color=red,
    mark=diamond,
    ]
    coordinates {
    (1,1.7006)(2,1.8623)(3,1.9632)(4,1.9799)
    %(10000,2.96)(19990,2.9)
    };
    \addlegendentry{\footnotesize CNS}
\addplot[
    color=blue,
    mark=square,
    ]
    coordinates {
    (1,1.3152)(2,1.8020)(3,2.0751)(4,2.2109)(5,2.2350)
   
    %(10000,2.9)(19990,2.5)
    };
    \addlegendentry{\footnotesize IC}
\addplot[
    color=green,
    mark=triangle,
    ]
    coordinates {
    
    (1,1.2761)(2,1.7550)(3,2.0116)(4,2.1291)(5,2.1881)(6,2.1957)
    %(15000,5.4)
    };
    \addlegendentry{\footnotesize SI}
\end{axis}
\end{tikzpicture}
% \includegraphics{}
%\caption{\tiny Jazz dataset}%Distribution of average distance with respect to number of iterations on Les Mis{\'e}rables dataset}
%\label{avgdiistancelesmiserables}
\end{subfigure}
 \hspace{0.12cm}
     \begin{subfigure}[b]{0.35\linewidth}
\centering
\usetikzlibrary{quotes}
\begin{tikzpicture}[scale=0.499]
\begin{axis}[
    title={Polblogs},%Iterations Vs Average Distance},
    xlabel={Iterations},
    ylabel={Average Distance},
    xmin=1, xmax=10,
    ymin=0, ymax=3,
    xtick={1,2, 3, 4,5,6,7,8,9,10},
    ytick={1.0,1.25,1.5,1.75,2.0,2.25,2.50,3.0},
    legend pos=south east,
    ymajorgrids=true,
    grid style=dashed,
]

\addplot[
    color=red,
    mark=diamond,
    ]
    coordinates {
    (1,1.7002)(2,2.2894)(3,2.5929)(4,2.6517)
    %(10000,2.96)(19990,2.9)
    };
    \addlegendentry{\footnotesize CNS}
\addplot[
    color=blue,
    mark=square,
    ]
    coordinates {
    (1,1.4)(2,2.0555)(3,2.6077)(4,2.7203)(5,2.7329)(6,2.7375)
   
    %(10000,2.9)(19990,2.5)
    };
    \addlegendentry{\footnotesize IC}
\addplot[
    color=green,
    mark=triangle,
    ]
    coordinates {
    
    (1,1.1666)(2,1.7644)(3,2.2443)(4,2.5886)(5,2.6761)(6,2.7042)(7,2.7211)(8,2.7314)(9,2.7329)(10,2.7375)
    %(15000,5.4)
    };
    \addlegendentry{\footnotesize SI}
\end{axis}
\end{tikzpicture}
% \includegraphics{}
%\caption{\tiny Polblogs dataset}%Distribution of average distance with respect to number of iterations on Les Mis{\'e}rables dataset}
%\label{avgdiiistancelesmiserables}
\end{subfigure} 
\caption{ Average distance within diffusion horizon per iteration.}% on real-world datasets. Higher scores indicate maximum information spread per iteration. }
\label{fig5}    
\end{figure}
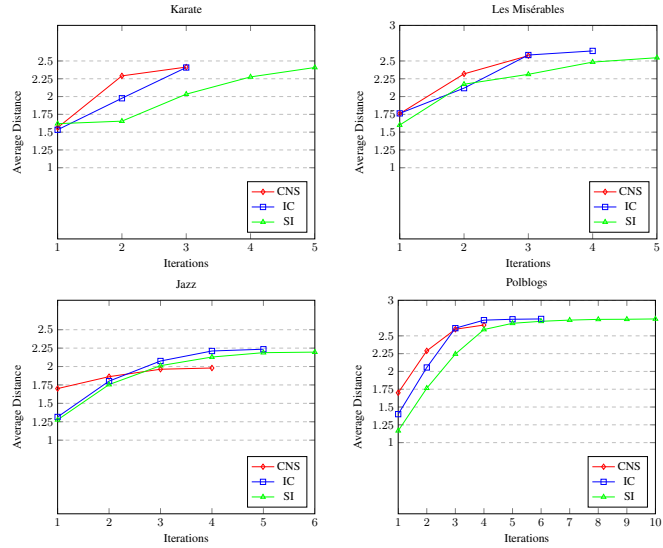

%///////////////////////////////////////////////////////////////////////////////%

\usetikzlibrary{positioning}
\tikzset{main node/.style={circle,fill=white,draw,minimum size=.2cm,inner sep=0pt},}
\begin{figure}[!h]
\centering
\begin{subfigure}[b]{0.35\linewidth}
\centering
\usetikzlibrary{quotes}
\begin{tikzpicture}[scale=0.4990]
\begin{axis}[
    title={Karate},%Iterations Vs Density},
    xlabel={Iterations},
    ylabel={Density},
    xmin=1, xmax=5,
    ymin=0, ymax=0.7,
    xtick={1,2, 3, 4,5},
    ytick={0.1,0.15,0.20,0.25,0.30,0.35,0.40,0.45,0.50},
    legend pos=north east,
    ymajorgrids=true,
    grid style=dashed,
]

\addplot[
    color=red,
    mark=diamond,
    ]
    coordinates {
    (1,0.4363)(2,0.1737)(3,0.1439)
    %(10000,2.96)(19990,2.9)
    };
    \addlegendentry{\footnotesize CNS}
\addplot[
    color=blue,
    mark=square,
    ]
    coordinates {
   (1,0.4666)(2,0.2094)(3,0.1390)
   
    %(10000,2.9)(19990,2.5)
    };
    \addlegendentry{\footnotesize IC}
\addplot[
    color=green,
    mark=triangle,
    ]
    coordinates {
    
    (1,0.3809)(2,0.4181)(3,0.2285)(4,0.1609)(5,0.1390)
    %(15000,5.4)
    };
    \addlegendentry{\footnotesize SI}
\end{axis}
\end{tikzpicture}
%\caption{\tiny Karate dataset }%Distribution of diameter with respect to number of iterations on}
%\label{denkarate}
    \end{subfigure}
     \hspace{0.12cm}
     \begin{subfigure}[b]{0.35\linewidth}
\centering
\usetikzlibrary{quotes}
\begin{tikzpicture}[scale=0.499]
\begin{axis}[
    title={Les Mis{\'e}rables},%Iterations Vs Density},
    xlabel={Iterations},
    ylabel={Density},
    xmin=1, xmax=5,
    ymin=0, ymax=1.0,
    xtick={1,2, 3, 4,5},
    ytick={0.1,0.20,0.30,0.4,0.5,0.6,0.7,0.8,0.9,1.0},
    legend pos=north west,
    ymajorgrids=true,
    grid style=dashed,
]

\addplot[
    color=red,
    mark=diamond,
    ]
    coordinates {
    (1,0.2363)(2,0.1207)(3,0.0943)
    %(10000,2.96)(19990,2.9)
    };
    \addlegendentry{\footnotesize CNS}
\addplot[
    color=blue,
    mark=square,
    ]
    coordinates {
   (1,0.2363)(2,0.1257)(3,0.0908)(4,0.8680)
   
    %(10000,2.9)(19990,2.5)
    };
    \addlegendentry{\footnotesize IC}
\addplot[
    color=green,
    mark=triangle,
    ]
    coordinates {
    
    (1,0.4)(2,0.1298)(3,0.1207)(4,0.1091)(5,0.1010)
    %(15000,5.4)
    };
    \addlegendentry{\footnotesize SI}
\end{axis}
\end{tikzpicture}
% \includegraphics{}
%\caption{\tiny Les Mis{\'e}rables dataset}%Distribution of diameter with respect to number of iterations on Les Mis{\'e}rables dataset}
%\label{denLESMISERABLES}
\end{subfigure} 
   \hspace{0.12cm}
     \begin{subfigure}[b]{0.35\linewidth}
\centering
\usetikzlibrary{quotes}
\begin{tikzpicture}[scale=0.4990]
\begin{axis}[
    title={Jazz},%Iterations Vs Density},
    xlabel={Iterations},
    ylabel={Density},
    xmin=1, xmax=6,
    ymin=0, ymax=1.0,
    xtick={1,2, 3, 4,5,6},
    ytick={0.10,0.20,0.3,0.4,0.5,0.6,0.7,0.8},
    legend pos=north east,
    ymajorgrids=true,
    grid style=dashed,
]

\addplot[
    color=red,
    mark=diamond,
    ]
    coordinates {
    (1,0.3212)(2,0.2259)(3,0.2041)(4,0.2014)
    %(10000,2.96)(19990,2.9)
    };
    \addlegendentry{\footnotesize CNS}
\addplot[
    color=blue,
    mark=square,
    ]
    coordinates {
   (1,0.6847)(2,0.2485)(3,0.1594)(4,0.1419)(5,0.1405)
   
    %(10000,2.9)(19990,2.5)
    };
    \addlegendentry{\footnotesize IC}
\addplot[
    color=green,
    mark=triangle,
    ]
    coordinates {
    
    (1,0.7238)(2,0.2846)(3,0.1761)(4,0.1532)(5,0.1447)(6,0.1433)
    %(15000,5.4)
    };
    \addlegendentry{\footnotesize SI}
\end{axis}
\end{tikzpicture}
% \includegraphics{}
%\caption{\tiny Jazz dataset}%Distribution of diameter with respect to number of iterations on Les Mis{\'e}rables dataset}
%\label{denjazz}
\end{subfigure} 
   \hspace{0.12cm}
     \begin{subfigure}[b]{0.35\linewidth}
\centering
\usetikzlibrary{quotes}
\begin{tikzpicture}[scale=0.4990]
\begin{axis}[
    title={ Polblogs},%Iterations Vs Density},
    xlabel={Iterations},
    ylabel={Density},
    xmin=1, xmax=10,
    ymin=0, ymax=1.0,
    xtick={1,2, 3, 4,5,6,7,8,9,10},
    ytick={0.1,0.2,0.3,0.4,0.5,0.6,0.7,0.8,0.9},
    legend pos=north east,
    ymajorgrids=true,
    grid style=dashed,
]

\addplot[
    color=red,
    mark=diamond,
    ]
    coordinates {
    (1,0.2997)(2,0.0514)(3,0.0276)(4,0.0260)
    %(10000,2.96)(19990,2.9)
    };
    \addlegendentry{\footnotesize CNS}
\addplot[
    color=blue,
    mark=square,
    ]
    coordinates {
   (1,0.6)(2,0.09670)(3,0.0255)(4,0.0225)(5,0.0224)(6,0.0224)
   
    %(10000,2.9)(19990,2.5)
    };
    \addlegendentry{\footnotesize IC}
\addplot[
    color=green,
    mark=triangle,
    ]
    coordinates {
    
    (1,0.833)(2,0.2401)(3,0.0675)(4,0.0275)(5,0.0241)(6,0.0233)(7,0.2279)(8,0.0225)(9,0.0224)(10,0.0224)
    %(15000,5.4)
    };
    \addlegendentry{\footnotesize SI}
\end{axis}
\end{tikzpicture}
% \includegraphics{}
%\caption{\tiny Polblogs dataset}%Distribution of diameter with respect to number of iterations on Les Mis{\'e}rables dataset}
%\label{densitypol}
\end{subfigure} 
\caption{Density within diffusion horizon per iteration.}% Higher density score when fraction of nodes affected per iteration is less  indicates low information spread per iteration. }
\label{fig6}    
\end{figure}
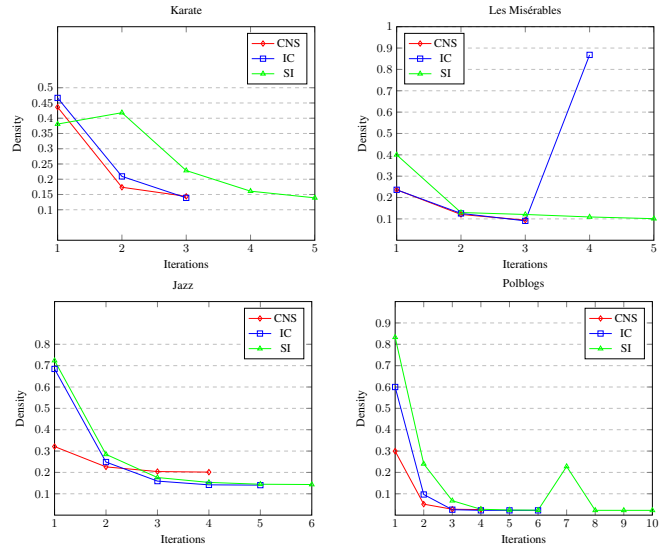

The comparative analysis of representative diffusion models with CNS algorithm is conducted considering diffusion speed and diffusion outspread. In this context, diffusion speed refers to `how fast the dispersion occurs'. Whereas, diffusion outspread indicates `how widely information propagates'. For evaluation of diffusion speed, we have examined the number of iterations required to reach steady state and fraction of nodes covered per iteration. Furthermore, to evaluate diffusion outspread, we have utilized graph properties such as, density, diameter, average distance, average degree. Comparative graphical results based on diffusion speed and diffusion outspread obtained by incorporation of CNS algorithm and representative information diffusion models on real-world datasets such as Karate, Les Mis{\'e}rables, Jazz and Polblogs obtained from SNAP~\cite{snap} repository are presented in this section. The details about these datasets are listed in Table~\ref{table1}. It is to be mentioned here that all the results presented here by incorporation of CNS algorithm, IC model and SI model assumes a common source/ infected node to initiate the diffusion process to maintain uniformity of our analysis.

\subsection{Result Analysis}
\label{ra}

The graphical results presented in Fig.~\ref{fig2} and Fig.~\ref{fig3} is related to diffusion speed. Here,  Fig.~\ref{fig2} refers to the total number of iterations taken to complete diffusion process and Fig.~\ref{fig3} refers to the fraction of nodes covered in each iteration. Less number of total iterations and maximum fraction of node coverage per iteration is expected for faster diffusion. As can be seen from Fig.~\ref{fig2}, SI model takes maximum number of iterations to complete diffusion process for all the representative datasets. Therefore, CNS algorithm is definitely better than SI model. Next, we need to compare the performance of CNS algorithm and IC model. From Fig.~\ref{fig2}, it is observed that CNS algorithm and IC model gives almost similar performance in terms of total number of iterations in small dataset such as Karate, but the difference in their performance can be identified in larger dataset as shown by Polblogs. As CNS algorithm takes least number of iterations as compared to IC model in large datasets, therefore it is inferred that CNS algorithm is better than IC model in terms of total number of iterations taken to complete diffusion process. Next, considering Fig.~\ref{fig3}, the half lines indicate that the respective information diffusion model/ algorithm completes in less number of iterations as compared to the SI model for all the representative datasets. Clearly, SI model covers least fraction of nodes per iteration. So, CNS algorithm is certainly better than SI model in terms of fraction of nodes covered per iteration. Additionally, it can be observed that CNS algorithm completes the diffusion process in three to four iterations for all the datasets and it covers maximum fraction of nodes in the first two iterations itself as compared to IC model. In particular, if we consider Polblogs dataset, it is clearly visible that CNS algorithm covers maximum fraction of nodes in the first two iterations in comparison to IC model. As CNS algorithm gives best performance in comparison to IC model and SI model based on total number of iterations required to complete diffusion process and fraction of nodes covered per iteration, therefore, CNS algorithm is better than representative diffusion models in terms of diffusion speed.

We perform a comparative analysis of CNS algorithm and representative information diffusion models in terms of diffusion horizon. In this context, diffusion horizon refers to the area covered by the diffusion process and this is measured per iteration. It is examined  to determine the diffusion outspread. The results shown in Fig.~\ref{fig4}, Fig.~\ref{fig5}, Fig.~\ref{fig6} and Fig.~\ref{fig7} is related to diffusion horizon.  Fig.~\ref{fig4} shows the diameter of diffusion horizon per iteration. Higher  diameter of diffusion horizon indicates maximum eccentricity from all nodes and hence, larger diffusion outspread. First of all, it is clearly visible from the subfigures of Fig.~\ref{fig4} that SI model gives least diameter of diffusion horizon in all the iterations for all the representative datasets. Therefore, CNS algorithm is certainly better than SI model in terms of diameter of diffusion horizon per iteration. Next, comparing CNS algorithm and IC model, both of these approaches gives almost similar performance for small datasets. But if we consider large dataset such as Polblogs, it can be seen that CNS algorithm gives maximum diameter of diffusion horizon per iteration compared to IC model. Therefore, CNS algorithm is better than the representative diffusion models in terms of diameter of diffusion horizon per iteration. Next,  Fig.~\ref{fig5} is used to evaluate average distance within diffusion horizon. Higher average distance covered per iteration indicates larger average shortest path distance and hence, wider diffusion outspread. Similar to diameter of diffusion horizon results, the performance of SI model is poor in terms of average distance within diffusion horizon. Whereas, comparison of CNS algorithm and IC model indicates that CNS algorithm covers maximum average distance in first two iterations as compared to IC model and it results in faster diffusion speed also. Therefore, CNS algorithm excels in performance in comparison to the representative diffusion models in terms of average distance within diffusion horizon. Next, Fig.~\ref{fig6} represents the density within diffusion horizon per iteration. Density is defined by $2m/n(n-1)$, where $m$ indicates number of edges and $n$ indicates number of nodes. It is used to measure the portion of potential connections in a network that are actual connections. It is expected that density decreases with increase in number of iterations for wider diffusion outspread because fraction of nodes covered in the initial iterations is expected to be high. As fraction of nodes covered per iteration is maximum for CNS algorithm, so the performance of CNS algorithm is best compared to SI model and IC model in terms of density within diffusion horizon. Fig.~\ref{fig7} indicates average degree of nodes within diffusion horizon per iteration. The criteria for a faster and wider diffusion is to target the maximum degree nodes first and hence, average degree is expected to rise and then fall with increase in number of iterations. As can be seen from Fig.~\ref{fig7}, CNS algorithm gives maximum average degree in the first two iterations for all the datasets compared to IC model and SI model which indicates best diffusion outspread of CNS algorithm in terms of average degree of nodes within diffusion horizon per iteration. Therefore, from these results, it is concluded that CNS algorithm gives best performance than representative diffusion models in terms of diffusion speed and diffusion outspread.

%%%%%%%%%%%%%%%%%%%%%%%%%%%%%%%%%%%%%%%%%%%%%%%%%%%%%%%%%%%%%%%%%%%

\section{Conclusion}
\label{conclusion}

\usetikzlibrary{positioning}
\tikzset{main node/.style={circle,fill=white,draw,minimum size=.2cm,inner sep=0pt},}
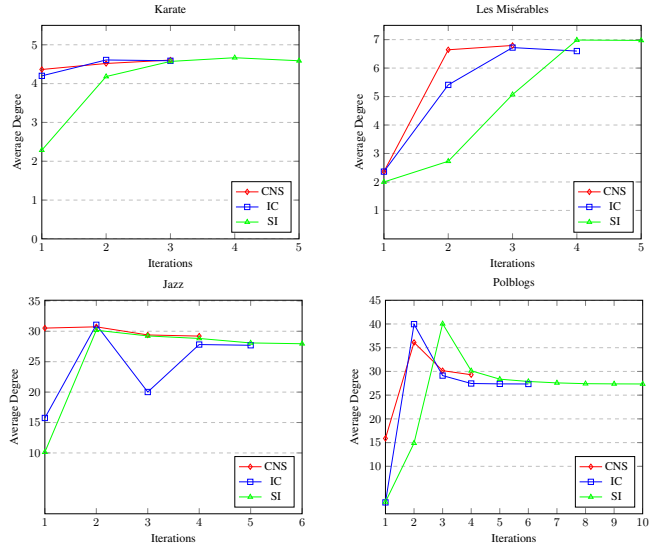
\begin{figure}[t]
\centering
\begin{subfigure}[b]{0.35\linewidth}
\centering
\usetikzlibrary{quotes}
\begin{tikzpicture}[scale=0.4990]
\begin{axis}[
    title={Karate},%Iterations Vs Average Degree},
    xlabel={Iterations},
    ylabel={Average Degree},
    xmin=1, xmax=5,
    ymin=0, ymax=5.5,
    xtick={1,2, 3, 4,5},
    ytick={1.0,2.0,3.0,4,0,5.0},
    legend pos=south east,
    ymajorgrids=true,
    grid style=dashed,
]

\addplot[
    color=red,
    mark=diamond,
    ]
    coordinates {
    (1,4.3636)(2,4.5185)(3,4.6060)
    %(10000,2.96)(19990,2.9)
    };
    \addlegendentry{\footnotesize CNS}
\addplot[
    color=blue,
    mark=square,
    ]
    coordinates {
   (1,4.2)(2,4.6086)(3,4.5882)
   
    %(10000,2.9)(19990,2.5)
    };
    \addlegendentry{\footnotesize IC}
\addplot[
    color=green,
    mark=triangle,
    ]
    coordinates {
    
    (1,2.2857)(2,4.1818)(3,4.5714)(4,4.6666)(5,4.5882)
    %(15000,5.4)
    };
    \addlegendentry{\footnotesize SI}
\end{axis}
\end{tikzpicture}
%\caption{\tiny Karate dataset }%Distribution of diameter with respect to number of iterations on}
%\label{avgdegkarate}
    \end{subfigure}
     \hspace{0.12cm}
     \begin{subfigure}[b]{0.35\linewidth}
\centering
\usetikzlibrary{quotes}
\begin{tikzpicture}[scale=0.499]
\begin{axis}[
    title={Les Mis{\'e}rables},%Iterations Vs Average Degree},
    xlabel={Iterations},
    ylabel={Average Degree},
    xmin=1, xmax=5,
    ymin=0, ymax=7.5,
    xtick={1,2, 3, 4,5},
    ytick={1,2,3,4,5,6,7},
    legend pos=south east,
    ymajorgrids=true,
    grid style=dashed,
]

\addplot[
    color=red,
    mark=diamond,
    ]
    coordinates {
    (1,2.3636)(2,6.6428)(3,6.7945)
    %(10000,2.96)(19990,2.9)
    };
    \addlegendentry{\footnotesize CNS}
\addplot[
    color=blue,
    mark=square,
    ]
    coordinates {
   (1,2.3636)(2,5.4090)(3,6.72)(4,6.5974)
   
    %(10000,2.9)(19990,2.5)
    };
    \addlegendentry{\footnotesize IC}
\addplot[
    color=green,
    mark=triangle,
    ]
    coordinates {
    
    (1,2.0)(2,2.7272)(3,5.0697)(4,6.9846)(5,6.9714)
    %(15000,5.4)
    };
    \addlegendentry{\footnotesize SI}
\end{axis}
\end{tikzpicture}
% \includegraphics{}
%\caption{\tiny Les Mis{\'e}rables dataset}%Distribution of diameter with respect to number of iterations on Les Mis{\'e}rables dataset}
%\label{avgdegLESMISERABLES}
\end{subfigure} 
   \hspace{0.12cm}
     \begin{subfigure}[b]{0.35\linewidth}
\centering
\usetikzlibrary{quotes}
\begin{tikzpicture}[scale=0.4990]
\begin{axis}[
    title={Jazz}, %Iterations Vs Average Degree},
    xlabel={Iterations},
    ylabel={Average Degree},
    xmin=1, xmax=6,
    ymin=0, ymax=35.1,
    xtick={1,2, 3, 4,5,6},
    ytick={10,15,20,25,30,35},
    legend pos=south east,
    ymajorgrids=true,
    grid style=dashed,
]

\addplot[
    color=red,
    mark=diamond,
    ]
    coordinates {
    (1,30.5208)(2,30.7299)(3,29.3931)(4,29.2054)
    %(10000,2.96)(19990,2.9)
    };
    \addlegendentry{\footnotesize CNS}
\addplot[
    color=blue,
    mark=square,
    ]
    coordinates {
   (1,15.75)(2,31.0634)(3,20.0163)(4,27.8172)(5,27.6969)
   
    %(10000,2.9)(19990,2.5)
    };
    \addlegendentry{\footnotesize IC}
\addplot[
    color=green,
    mark=triangle,
    ]
    coordinates {
    
    (1,10.1333)(2,30.1682)(3,29.2335)(4,28.8148)(5,28.0820)(6,27.9489)
    %(15000,5.4)
    };
    \addlegendentry{\footnotesize SI}
\end{axis}
\end{tikzpicture}
% \includegraphics{}
%\caption{\tiny Jazz dataset}%Distribution of diameter with respect to number of iterations on Les Mis{\'e}rables dataset}
%\label{avgdegjazz}
\end{subfigure} 
   \hspace{0.12cm}
     \begin{subfigure}[b]{0.35\linewidth}
\centering
\usetikzlibrary{quotes}
\begin{tikzpicture}[scale=0.4990]
\begin{axis}[
    title={Polblogs},%Iterations Vs Average Degree},
    xlabel={Iterations},
    ylabel={Average Degree},
    xmin=1, xmax=10,
    ymin=0, ymax=45,
    xtick={1,2, 3, 4,5,6,7,8,9,10},
    ytick={10,15,20,25,30,35,40,45},
    legend pos=south east,
    ymajorgrids=true,
    grid style=dashed,
]

\addplot[
    color=red,
    mark=diamond,
    ]
    coordinates {
    (1,15.8888)(2,36.1109)(3,30.1596)(4,29.3019)
    %(10000,2.96)(19990,2.9)
    };
    \addlegendentry{\footnotesize CNS}
\addplot[
    color=blue,
    mark=square,
    ]
    coordinates {
   (1,2.4)(2,39.9613)(3,29.1210)(4,27.4576)(5,27.3759)(6,27.3551)
   
    %(10000,2.9)(19990,2.5)
    };
    \addlegendentry{\footnotesize IC}
\addplot[
    color=green,
    mark=triangle,
    ]
    coordinates {
     (1,2.5)(2,14.8888)(3,40.0437)(4,30.1352)(5,28.3696)(6,27.8813)(7,27.5854)(8,27.4175)(9,27.3759)(10,27.3551)
    
    %(15000,5.4)
    };
    \addlegendentry{\footnotesize SI}
\end{axis}
\end{tikzpicture}

\end{subfigure} 
\caption{\footnotesize Average degree of nodes within diffusion horizon per iteration.}% Higher scores indicate maximum information spread per iteration. }
\label{fig7}    
\end{figure}

In this paper, we developed a network property based information diffusion algorithm called CNS considering both dense and sparse networks. It utilizes common neighborhood information to compute tie strength score. It is based on the concept that strong ties reside within densely connected nodes. Extensive experiments on several real-world datasets show that CNS algorithm achieves best diffusion speed and diffusion outspread among IC model and SI model. Therefore, it is inferred that network property plays a significant role in the dynamics of information diffusion. In future, we will deploy social aspects in combination with network property aspects to design information diffusion method and examine it's significance.

\end{document}